\documentclass[11pt]{article}
\usepackage{graphicx,amssymb,amsmath}
\def\etal{{et~al.}}
\def\Ne{\ensuremath{N_{\rm e}}}
\def\Nreal{\ensuremath{N_\mathrm{real}}}
\def\Nbg{\ensuremath{N_\mathrm{bg}}}

\def\Pbc{\ensuremath{P_\mathrm{bc}}}
\def\PN{\ensuremath{P_N}}
\def\Ep{\ensuremath{E_\mathrm{p}}}
\def\EFe{\ensuremath{E_\mathrm{Fe}}}

\newcommand{\art}[5]{#1, #2 \textbf{#3}, #5 (#4).}
\newcommand{\artx}[6]{#1, #2 \textbf{#3}, #5 (#4); \texttt{arXiv:#6}.}
\newcommand{\prep}[3]{#1, \textit{#2}, \texttt{arXiv:#3}.}
\newcommand{\proc}[4]{#1, \textit{#2} (#3),        p.~#4.}
\newcommand{\book}[3]{#1, \textit{#2} (#3).}
\begin{document}
\title{\bfseries
Three Regions of Excessive Flux of PeV Cosmic Rays%
\footnote{%
The preprint fixes errors that appeared in the English version
of the article published in Bulletin of the Russian Academy of Sciences.
Physics, 2011, Vol.~75, No.~3, pp.~342--346. Original Russian text:
Izvestiya Rossiiskoi Akademii Nauk. Seriya Fizicheskaya, 2011, Vol.~75,
No.~3, pp.~371--375.
}
}

\author{%
G. V. Kulikov, M. Yu.\ Zotov\\[1mm]
D. V. Skobeltsyn Institute of Nuclear Physics,\\
M. V. Lomonosov Moscow State University,\\
Moscow 119992, Russia\\
\texttt{\{kulikov,zotov\}@eas.sinp.msu.ru}
}

\date{}
\maketitle

\begin{abstract}
Three regions of excessive flux of cosmic rays with energies of the order
of PeV are found in the experimental data of the EAS MSU array at a
confidence level greater than $4\sigma$. For two of them, there are
similar regions in the experimental data of the EAS--1000 Prototype
array. One of the interesting features of the regions is the absence of
supernova remnants in their vicinities, traditionally considered as the
main sources of Galactic cosmic rays, but the presence of isolated pulsars,
some of which are able to accelerate heavy nuclei up to energies close to PeV.
In our opinion, this favors the assumption that isolated pulsars are able
to contribute to the flux of Galactic cosmic rays more than is usually
assumed.
\end{abstract}

\section{Introduction}

The problem of the origin of cosmic rays (CRs) in the PeV energy range
remains open for several decades. The model of diffusive shock
acceleration at the outer front of expanding supernova remnants (SNRs)
has become the most widely spread one, see, e.g.,~[1], though it has not
been experimentally confirmed yet. Along with supernova remnants, other
possible sources of PeV CRs have been considered. In particular, shortly
after pulsars were discovered and identified with neutron stars, it was
demonstrated that they can be effective accelerators of CRs~[2].
Interest to pulsars as possible sources of CRs with energies
$\gtrsim10^{14}$~eV and up to ultrahigh energies did not vanish [3--7].
Models considered included acceleration both in pulsar wind nebulae
(e.g., the Crab nebula) and by isolated pulsars. It was demonstrated in
particular that the Geminga pulsar is a possible candidate for being the
single source of the knee at 3~PeV~[5]. There is no surprise then that
one of the areas of investigations in the field are attempts to find
correlation between arrival directions of CRs and coordinates of their
possible astrophysical sources, including SNRs, pulsars, and some other.

In our previous works dedicated to the analysis of arrival directions of
extensive air showers (EAS) registered with the EAS MSU and EAS--1000
Prototype (below, ``PRO--1000'') arrays, we have already presented a
number of regions of excessive flux (REFs) of PeV cosmic rays [8--11]. It
was shown that  not all of them could be matched with Galactic supernova
remnants. However, a substantial part of REFs could be matched with
isolated pulsars in whose vicinities nebulae are not observed. In this
work, we consider three regions selected in the EAS MSU array data at a
level of confidence greater than $4\sigma$. None of these could be
matched with Galactic supernova remnants, but there are isolated pulsars
inside or in close vicinities of these regions, some of which are able to
accelerate heavy nuclei up to energies close to PeV. For two of these
regions, there are similar regions in the PRO--1000 experimental data.

\section{Experimental Data and Method of The Investigation}

As in our previous studies of EAS arrival directions, the main data sets
consist of 513,602 showers registered with the EAS MSU array in
1984--1990 and 1,342,340 showers of the PRO--1000 array registered in
1997--1999.  A description of the arrays can be found in~[12]
and~[13] respectively. All EAS selected for the analysis satisfy a number
of quality criteria and have zenith angles $\theta < 45.7^\circ$. Showers
registered with the two arrays have different number of charged
particles~$\Ne$ in a typical event.  For the EAS MSU array, the median
value of~$\Ne$ is of the order of $1.6\times10^5$, while that for the
PRO--1000 array equals $3.7\times10^4$. According to the modern models of
hadronic interactions and data on the chemical composition of cosmic
rays, the given values of~\Ne\ correspond to the energies of a primary
proton $E_0\approx1.7\times10^{15}$~eV and
$E_0\approx4.6\times10^{14}$~eV respectively, with an error of about
10\%--20\%. The EAS MSU array allowed one to determine arrival directions
of EAS with a better accuracy than the PRO--1000 array but the mean error
is estimated to be of the order of~$3^\circ$ in both cases.

The investigation is based on the method by Alexandreas \etal, which was
developed for the analysis of arrival directions of EAS registered with
the CYGNUS array~[14] and since then has been used multiple times for the
analysis of results of other experiments on CRs. The idea of the method
is as follows. To every shower in the experimental data set, arrival time
of another shower is assigned in a pseudo-random way. After this, new
equatorial coordinates $(\alpha,~\delta)$ are calculated for the
``mixed'' data set thus providing a ``mixed'' map of arrival directions. 
The mixed map differs from the original one but has the same distribution
in declination~$\delta$. In order to compare both maps, one divides them
into sufficiently small ``basic'' cells. A measure of difference between
any two regions (cells) of the two maps located within the same
boundaries is defined as \[ S = (\Nreal - \Nbg)/\sqrt{\Nbg}, \] where
\Nreal\ and \Nbg\ are the number of EAS inside a cell in the real and
``mixed'' (background) maps correspondingly. The mixing of the real map
is performed multiple times, and the mixed maps are averaged then in
order to reduce the dependence of the result on the choice of arrival
times.  The method is based on an assumption that the resulting mean
``background'' map has most of the properties of an isotropic background,
and presents the distribution of arrival directions of cosmic rays that
would be registered with the array in case there is no anisotropy. Thus,
deviations of the real map from the background one may be assigned to a
kind of anisotropy of arrival directions of EAS registered with the array.
As a rule, selection of regions of excessive flux is performed basing on
the condition $S>3$.

As in~[15], the basic cells have the size $0.5^\circ\times0.5^\circ$ and
the number of cycles of mixing equals 10,000. Due to this, the difference
in the number of showers in basic cells of any two sequentially averaged maps
by the end of the mixing cycle
is of the order or (1--2)$\times10^{-3}$ for both data sets.

The procedure of constructing cells (regions) for comparing the number of EAS
falling into them is as follows. Adjacent basic strips in~$\delta$ (each
$0.5^\circ$ wide) were joined into strips of width
$\Delta\delta=3^\circ\dots30^\circ$ with the step equal to~$0.5^\circ$.
Each wide strip was then divided into adjacent cells of some fixed
width~$\Delta\alpha$. After this, we calculated the number of EAS inside
each of these cells for both experimental (\Nreal) and background (\Nbg)
maps. For each pair of cells, $S$ was calculated then. A cell was
considered as a possible cell of excessive flux (CEF) if $S>3$. Every
wide strip was divided into cells with a shift of $0.5^\circ$
in the angular distance in~$\alpha$
until all possible locations on the grid were covered. Contrary to~[11],
we did not restrict the search to cells such that
$\Delta\alpha=\Delta\delta/\cos\bar{\delta}$, where~$\bar\delta$ is the
mean value of~$\delta$ for the current strip, and $\Delta\alpha$ is
rounded to the nearest half-integer number, but allowed $\Delta\alpha$ to
deviate from that value in such a way that the area of the resulting
cells deviated from the original one by~1/6. As a result, additional
cells were considered.  For example, besides cells of the size
$\Delta\alpha\times\Delta\delta=3^\circ\times3^\circ$, we also considered
cells of the sizes $2.5^\circ\times3^\circ$ and $3.5^\circ\times3^\circ$.
Finally, only cells with more than 100 real EAS inside were analyzed.

The method of Alexandreas et~al. does not provide a direct answer to the
question about the chance probability of appearance of a CEF.
One is expected to calculate the corresponding probability basing on the
value of~$S$, which is an estimate of the standard deviation of a sample
and thus acts as a significance level.
It is implicitly assumed by this that the deviation of~\Nreal\ from~\Nbg\
has a Gaussian distribution.
Hence, the chance probability of a CEF to appear is estimated to be less
than $1-0.9973$ providing it was selected at significance level $S>3$.

In order to estimate the chance probability of appearance of a CEF basing
on the number of EAS inside, we introduced the following simple method
based on the binomial distribution~[11]. Let a shower
axis getting inside the CEF be a success. The number of trials equals the
number of showers~$N$ in the data set under consideration, and an
estimate of success (for a fixed region) equals $\tilde p = \Nbg/N$,
where~\Nbg\ is the number of showers in the cell of the background map.
The assumption is based on the fact that in the method of Alexandreas
\etal, \Nbg\ is considered to be an expected number of showers in a cell.
Obviously, the chance probability of finding exactly~\Nreal\ EAS in a
cell (or region) equals
\[
	P(\nu=\Nreal) = C(N,\Nreal) \tilde{p}^{\Nreal} (1 - \tilde{p})^{N-\Nreal},
\]
where $\nu$ is a random variable equal to the number of successes in the
binomial model, and $C(N,\Nreal)$ is the corresponding binomial
coefficient. It is more interesting to consider a probability that there
are at most~\Nreal\ showers in the cell $\PN=P(\nu\le\Nreal)$. The
analysis performed and the data presented below demonstrate that values
of~\PN\ correlate well with the values of chance probabilities calculated
on the basis of significance level~$S$.

The fact that $S>3$ for a given cell does not imply it is valid for its
subcells. A situation such that $S<3$ for a number of adjacent cells but
$S>3$ for a cell (region) combined of them is possible. In order to
improve the robustness of selection of CEFs, we have tried a number of
additional quantities. One of the most efficient of them is the
probability $\Pbc=P(\xi<N_\mathrm{bc}^+)$, calculated from the following
binomial model. Let~$\xi$ be a random variable equal to the number of
basic cells of the given cell with an excess of EAS over the background
values, and $N_\mathrm{bc}^+$ is the corresponding experimental value.
The number of trials equals the number of basic cells in the CEF. It is
natural to assume the probability of success to be equal to~1/2. In the
results presented below, all CEFs satisfy a condition $\Pbc>0.9545$,
which corresponds to the significance level of $2\sigma$ for the Gaussian
distribution.  We thus reduced the chance that a CEF is selected solely
due to the non-uniformity of the EAS distribution w.r.t.~$\delta$.

\section{Main Results and Discussion}

In what follows, we consider REFs found in the EAS MSU array
experimental data under the conditions
\begin{equation}
S > 4,\quad \Pbc > 0.9545,\quad \Nreal > 100.
\end{equation}
Three regions satisfying~(1) were found, see Fig.~1 and Table~1. Each of
them consists of a number of partially overlapping cells, selected with
the algorithm described above. The probability that any of the three REFs
appears by chance is $\le6\times10^{-5}$. This estimate follows from the
values of~\PN\ shown in Table~1, as well as from the fact that $S>4$ for
all the REFs: the corresponding probability equals $\approx0.999937$ for
the significance level equal to~4. Regions~A and~B have ``partners'' in
the PRO--1000 data set. These are regions that satisfy conditions $S>3$,
$\Pbc>0.9545$, $\Nreal>100$ and partially overlap with those two regions
in the EAS MSU data set.

\begin{figure}[!ht]
\centerline{\includegraphics[width=.7\textwidth]{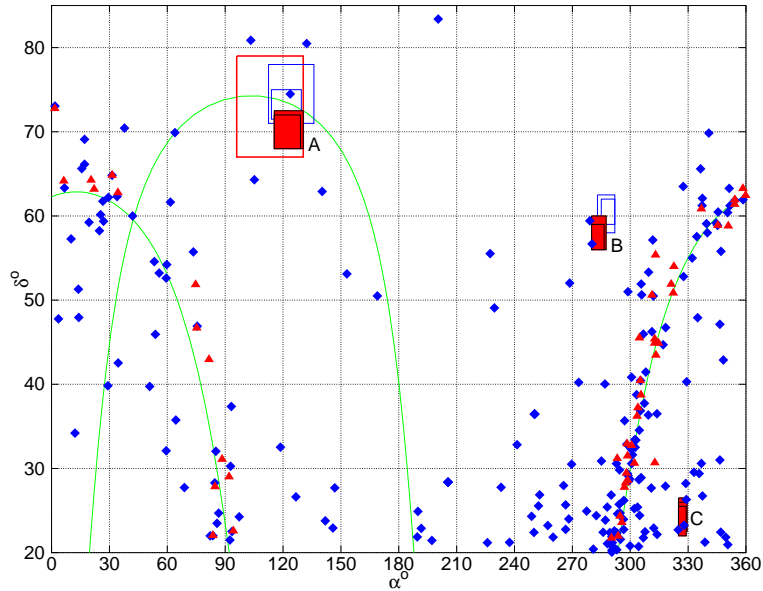}}
\vspace{-1pc}
\caption{Location of the regions of excessive flux in equatorial coordinates.
Filled with red are the three regions found in the EAS MUS data that
satisfy conditions~(1). The open red rectangle shows the boundaries of
a region found with the condition $\Pbc>0.9545$ omitted. Blue rectangles
show REFs found in the PRO--1000 data set that partially overlap
with the above REFs in the EAS MSU data and satisfy
conditions $S>3$, $\Pbc>0.9545$, $\Nreal>100$ (``partners'').
Filled red triangles show coordinates of the Galactic supernova remnants~\cite{Green}.
Filled blue diamonds show coordinates of the Galactic pulsars~\cite{ATNF}.
The curves at the left and right sides of the Figure show the Galactic plane,
the $\cap$-like curve shows the Supergalactic plane.}
\end{figure}

\begin{table}[!ht]
\label{tab:MSU}
\caption{Some parameters of the REFs in the EAS MSU data set that satisfy
conditions~(1) (A, B, C),
and their ``partners'' in the PRO--1000 data set (A$'$, B$'$).
Notation:
$\alpha$ and $\delta$ are the ranges of values of the corresponding coordinates;
\Nreal, \Nbg\ are the ranges of the experimental and background flux of EAS
in the cells that constitute the REF respectively,
$\max S$ and $\min\PN$ are the maximum value of the significance level~$S$
and the minimum value of~\PN\ for the cells that constitute the REF.
}
\begin{center}
\begin{tabular}{|c|c|c|c|c|c|c|}
\hline
REF & $\alpha,\,^\circ$ & $\delta,\,^\circ$ & \Nreal\ & \Nbg\ & $\max S$ & $\min\PN$ \\
\hline
A & 115.5\dots130.5 & 68\dots72.5 & 960\dots1498 & 841.9\dots1350.8 & 4.233461 & 0.999962 \\
A$'$& 112.5\dots136 & 71\dots78 & 2034\dots6406 & 1881.4\dots6146.9 & 3.852460 & 0.999474 \\
\hline
B & 280\dots287.5 & 56\dots60 & 537\dots1029 & 451.5\dots904.3 & 4.478172 & 0.999958 \\
B$'$& 283\dots292 & 58\dots62.5 & 1687\dots3473 & 1552.4\dots3283.8 & 3.556190 & 0.999494 \\
\hline
C & 325\dots329.5 & 22\dots26.5 & 108\dots152 & 67.34\dots108.9 & 4.954917 & 0.999939 \\
\hline
\end{tabular}
\end{center}
\end{table}

The median values of~\Ne\ for the regions A, B, and C are close to the
value obtained for the whole data set. The region that is found around
region~A when omitting the condition $\Pbc>0.9545$ lies whithin the
boundaries $\alpha=96^\circ\dots130.5^\circ$ and
$\delta=67^\circ\dots79^\circ$ and contains 7386 EAS. The median value
of~\Ne\ for this REF is almost equal to the value for the whole data set.
The minimum values of~\Pbc\ and~\PN\ for the cells forming this region
are equal to 0.831062 and 0.999962, respectively.

As can be seen from the Figure, there are currently no known Galactic
supernova remnants in the vicinity of the regions found.  Some other
possible accelerators of Galactic CRs, such as X-ray binaries and
OB-associations have not been found there either. There is a bright
open cluster NGC 2408 at the western boundary of Region~A, but the
distance from it to the Solar system is unknown, and it might be difficult
to obtain a reliable estimation of its contribution to the flux
of Galactic PeV CRs. On the other hand, there are isolated pulsars inside
or in the close vicinities of all the regions. In view of this, the
question arises as to whether these pulsars could accelerate charged
particles to the energies of the order of PeV?

To the best of our knowledge, there is currently no common model of
acceleration of charged particles in the vicinity of isolated pulsars.
To estimate the maximum energy of a particle accelerated near the light
cylinder of a pulsar, we employ an expression that can be written as
follows: $E_{\max} = 0.34 \, Z \, B \, \Omega^2$~eV,
where $Z$ is the charge number of the particle, $B$ is the strength of
the surface magnetic field,~G; $\Omega$ is the angular velocity of a
pulsar, rad~s$^{-1}$; and the radius of the pulsar equals $10^6$~cm~[3].
The expression was obtained under the assumption that most of the
magnetic field energy in the pulsar wind zone is transformed into the
kinetic energy of the particles, and the density of electron-positron
pairs does not exceed $10^{-5}$ of that of ions of iron.
Table~2 presents approximate values of maximum energies of protons and
iron nuclei accelerated by pulsars located in the vicinities of
regions~A, B, and C. It can be seen that in all three cases sufficiently
heavy nuclei can be accelerated up to energies close to PeV.

\begin{table}[!ht]
\label{tab:PSRs}
\caption{Some parameters of the pulsars located near REFs A, B, and C, and
estimates of the maximum energy of protons and iron nuclei. Notation:
$d$ is the distance from a pulsar to the Solar system,~kpc;
$P$ is the barycentric period of rotation of a pulsar,~s;
$B$ is the strength of the surface magnetic field,~G;
$\dot E$ is the rotation energy loss,~ergs/s;
$\max\Ep$ and $\max\EFe$ are the maximum energies of a proton and
an iron nucleus accelerated by a pulsar,~eV.}
\begin{center}
\begin{tabular}{|c|c|c|c|c|c|c|c|c|}
\hline
REF& PSR Name & $d,$ & $P,$  & Age,              & $B,$     & $\dot E,$ & $\max\Ep,$ & $\max\EFe,$ \\
 &            & kpc  &  s    & year              &  G       & ergs/s    &  eV        & eV \\
\hline
A& J0814+7429 & 0.33--0.43 & 1.292 & $1.22\cdot10^{8}$ & $4.72\cdot10^{11}$ & $3.08\cdot10^{30}$ & $4\cdot10^{12}$ & $1\cdot10^{14}$  \\
 & J0700+6418 & 0.48 & 0.196 & $4.52\cdot10^{9}$ & $1.17\cdot10^{10}$ & $3.61\cdot10^{30}$ & $4\cdot10^{12}$ & $1\cdot10^{14}$  \\
 & J0653+8051 & 3.37 & 1.214 & $5.07\cdot10^{6}$ & $2.17\cdot10^{12}$ & $8.37\cdot10^{31}$ & $2\cdot10^{13}$ & $5\cdot10^{14}$  \\
 & J0849+8028 & 3.38 & 1.602 & $5.69\cdot10^{7}$ & $8.56\cdot10^{11}$ & $4.28\cdot10^{30}$ & $4\cdot10^{12}$ & $1\cdot10^{14}$  \\
\hline
B& J1836+5925 &0.17--0.75& 0.173 & $1.80\cdot10^{6}$ & $5.20\cdot10^{11}$ & $1.16\cdot10^{34}$ & $2\cdot10^{14}$ & $6\cdot10^{15}$  \\
 & J1840+5640 & 1.70 & 1.653 & $1.75\cdot10^{7}$ & $1.59\cdot10^{12}$ & $1.31\cdot10^{31}$ & $8\cdot10^{12}$ & $2\cdot10^{14}$  \\
\hline
C& J2155+2813 & 5.06 & 1.609 & $2.78\cdot10^{7}$ & $1.23\cdot10^{12}$ & $8.68\cdot10^{30}$ & $6\cdot10^{12}$ & $2\cdot10^{14}$  \\
 & J2156+2618 & 4.71 & 0.498 & $5.56\cdot10^{8}$ & $8.51\cdot10^{10}$ & $4.53\cdot10^{30}$ & $5\cdot10^{12}$ & $1\cdot10^{14}$  \\
 & J2151+2315 & 1.42 & 0.594 & $1.33\cdot10^{7}$ & $6.56\cdot10^{11}$ & $1.34\cdot10^{32}$ & $2\cdot10^{13}$ & $6\cdot10^{14}$  \\
 & J2139+2242 & 4.71 & 1.084 & $1.21\cdot10^{7}$ & $1.26\cdot10^{12}$ & $4.41\cdot10^{31}$ & $1\cdot10^{13}$ & $4\cdot10^{14}$  \\
\hline
\end{tabular}
\end{center}
\end{table}

The distances to the pulsars from Table~2 exceed considerably the free
path of PeV neutrons, and the amount of electromagnetic showers is
insufficient for formation of REFs. In view of this, the question arises
of how to keep the arrival direction of (charged) CRs propagating through
the Galactic magnetic field.  We are currently not ready to suggest a
satisfactory model for answering this question, but we note at least two
models were suggested to explain the REFs of CRs with energies of
the order of 10~TeV registered with the Milagro gamma-ray
observatory~[18]. One of the models is based on CR acceleration in the
vicinity of the Geminga pulsar [19], the properties of which are similar
to those of the pulsar J1836+5925 located at the boundary of the B
region~[20]. The other one uses a specific magnetic field configuration
able to focus a flux of charged particle~[21].

To conclude, we have presented three regions of excessive flux of PeV CRs
selected from the EAS MSU experimental data at a significance level
$>4\sigma$ and satisfying several additional conditions. In the vicinity
of every region, there are isolated Galactic pulsars that are able to
accelerate sufficiently heavy nuclei to the required energies.
In our opinion, this does
not give enough ground to claim that the above REFs are formed due to
contribution from these pulsars but can witness in favor of the
assumption that (1) some isolated pulsars contribute to the flux of PeV
CRs more than is usually assumed and (2) there are
efficient mechanisms of focusing CR in the interstellar
space.

\bigskip

Only free, open source software was used for the investigation. In
particular, all calculations were performed with GNU Octave~[22]
running in Linux. This research has made use of the SIMBAD
database, operated at CDS, Strasbourg, France
(\texttt{http://simbad.u strasbg.fr/simbad}).

The research was partially supported by the Russian Foundation for
Fundamental Research grant No.~08-02-00540


\end{document}